\documentclass[10pt,conference]{IEEEtran}
\IEEEoverridecommandlockouts
\usepackage{float}
\usepackage{graphicx}
\usepackage{amsmath}
\usepackage{amsfonts}
\usepackage{mathtools}
\usepackage{dirtytalk}
\usepackage{listings}
\usepackage{tikz}
\usepackage{enumitem}
\usepackage{todonotes}
\usepackage{subfigure}
\usepackage{array}
\usepackage{multirow, multicol}
\usepackage{adjustbox}
\usepackage{makecell}
\usepackage{array}
\usepackage{multirow}
\usepackage{tcolorbox}
\usepackage{cite}
\usepackage{amsmath,amssymb,amsfonts}
\usepackage{algorithmic}
\usepackage{graphicx}
\usepackage{textcomp}
\usepackage{xcolor}
\usepackage{url} 
\usepackage{hyperref}
\usepackage{orcidlink}
\usepackage{booktabs, adjustbox}

\usetikzlibrary{fit}

\graphicspath{ {../Thesis/} }

\newcounter{findingctr}

\lstdefinestyle{bashstyle}{
  language=bash,
  basicstyle=\ttfamily\small,
  keywordstyle=\color{blue},
  commentstyle=\color{green},
  stringstyle=\color{red},
  showstringspaces=false,
  breaklines=true,
  backgroundcolor=\color{lightgray},
  frame=single,
  framerule=0.5pt,
  rulecolor=\color{gray}
}

\newcommand\copyrighttext{%
  \footnotesize © 2025 IEEE. Personal use of this material is permitted.
  Permission from IEEE must be obtained for all other uses, in any current or future
  media, including reprinting/republishing this material for advertising or promotional
  purposes, creating new collective works, for resale or redistribution to servers or
  lists, or reuse of any copyrighted component of this work in other works.
  DOI: 10.1109/GREENS66463.2025.00010. Available at:\\ \url{https://doi.ieeecomputersociety.org/10.1109/GREENS66463.2025.00010}}
\newcommand\copyrightnotice{%
\begin{tikzpicture}[remember picture,overlay]
\node[anchor=south,yshift=10pt] at (current page.south) {\fbox{\parbox{\dimexpr\textwidth-\fboxsep-\fboxrule\relax}{\copyrighttext}}};
\end{tikzpicture}%
}

\begin{document}

\title{An exploration of prompting LLMs to generate energy-efficient code}

\author{
\IEEEauthorblockN{Tom Cappendijk \orcidlink{0009-0006-1547-9079}}
\IEEEauthorblockA{\textit{Informatics Institute} \\
\textit{Universiteit van Amsterdam}\\
Amsterdam, The Netherlands \\
tom.cappendijk@student.uva.nl}
\and
\IEEEauthorblockN{Pepijn de Reus \orcidlink{0009-0008-0087-5970}}
\IEEEauthorblockA{\textit{Informatics Institute} \\
\textit{Universiteit van Amsterdam}\\
Amsterdam, The Netherlands \\
p.dereus@uva.nl}
\and
\IEEEauthorblockN{Ana Oprescu \orcidlink{0000-0001-6376-0750}}
\IEEEauthorblockA{\textit{Informatics Institute} \\
\textit{Universiteit van Amsterdam}\\
Amsterdam, The Netherlands \\
a.m.oprescu@uva.nl}
}

\maketitle

\begin{abstract}
The increasing electricity demands of personal computers, communication networks, and data centers contribute to higher atmospheric greenhouse gas emissions, which in turn lead to global warming and climate change. Therefore the energy consumption of code must be minimised. Large language models can generate code, so we study the influence of prompting for energy-efficient code by examining the energy consumption of the generated code. We use three different Python code problems of varying difficulty levels. Prompt modification is done by adding the sentence ``Give me an energy-optimised solution for this problem'' or by providing two Python coding best practices. The large language models used are Code Llama-70b, Code Llama-70b-Instruct, Code Llama-70b-Python, DeepSeek-Coder-33b-base, and DeepSeek-Coder-33b-instruct. We find a decrease in energy consumption for a specific combination of prompt optimisation, LLM, and Python code problem. However, no single optimisation prompt consistently decreases energy consumption for the same LLM across the different Python code problems.
\end{abstract}


\copyrightnotice

\section{Introduction}
The increase in atmospheric greenhouse gasses (GHG) emission has led to global warming~\cite{kweku2018greenhouse}. Zero emission of GHG is mandatory to stall climate change. While the European Union's Green Deal sets the zero-emission goal for 2050~\cite{EuropeanCouncil}, the electricity demand for IT infrastructure increases~\cite{the_shift_project_energy_2023}. Reducing the energy consumption of computing systems reduces the emission of GHG~\cite{radersma2022green}, and can, amongst others, be achieved by optimising the software used in these systems. This study focuses on software, in particular, code that is generated by Large Language Models (LLMs). The ability to generate correct code by a LLM is not positively correlated to the ability to generate efficient code~\cite{niu2024evaluating}. So for a user, it is important to judge the LLM not solely on the correctness of code but also on the code's energy consumption.

Lately, a growing number of studies examine the energy consumption and carbon emissions associated with LLMs~\cite{bender2021dangers, patterson2021carbon}. E.g. by investigating strategies to lower the energy consumption of these models during training\cite{mcdonald2022great, de2021hyperparameter}. However, during usage, a LLM requires computation as well. As such, another research branch focuses on the energy usage during inference and provides solutions to lower energy consumption~\cite{samsi2023words, wang2020energy}. To position our research within the domain of energy consumption of LLMs, we do not focus on energy consumption during training or inference. Instead, we focus on the energy consumption of the code produced by LLMs. Our work relates to very recent research which assesses the sustainability awareness of LLMs by reviewing the code they produce~\cite{vartziotis2024learn, cursaru2024controlled}, and differs in that we focus on the ability of LLMs to generate code with a lower energy consumption. We conduct experiments by prompting energy-efficient code solutions and collect the statistics, including energy consumption, of the code generated by LLMs.

To understand to what extent can prompts trigger a LLM to produce code that consumes less energy, we formulate the following two research questions:

\begin{enumerate}[label=(\arabic*)]
  \item \textbf{Which prompt optimisation leads to generating code that is consuming less energy for a given LLM?}\label{question:2.1}
  \item \textbf{To what extent does the code problem influence the impact of the prompt optimisation on energy consumption?}\label{question:2.2}
\end{enumerate}

We use the programming language Python and three difficulty levels. Prompts are modified by adding the sentence ``Give me an energy-optimised solution for this problem'' or by using two Python coding best green practices~\cite{koedijk2022finding}. We use Code Llama-70b, Code Llama-70b-Instruct, Code Llama-70b-Python~\cite{roziere2023code}, DeepSeek-Coder-33b-base, and DeepSeek-Coder-33b-instruct~\cite{guo2024deepseek} and find a decrease in energy consumption for a specific combination of prompt optimisation, LLM, and Python code problems. However, no single optimisation prompt consistently decreases energy consumption for the same LLM across the different Python code problems.

\textbf{Scope of this paper} We do not incorporate all the available LLMs. There is a continuous release of new models. Some of the LLMs, such as ChatGPT-3, are behind a paywall and fall out of our scope. We also did not set out to implement open-source LLMs. Our focus is on prompt modification, showing how prompts can trigger a LLM to produce code that consumes less energy. Additionally, fine-tuning the implementation of a specific LLM can be complex due to numerous parameters and training options. Delving into such fine-tuning is not within the scope of this study.

\section{Background and Related Work} \label{sec:background}
A language model is a model that assigns probabilities to upcoming words or sequences of words~\cite{jurafsky2024speech}. A LLM is the product of a pretrained language model, pretraining enables the LLM to gain knowledge about the language and its context based on large amounts of text. After pretraining, a LLM is fine-tuned for tasks such as questioning and answering.

\subsection{An Empirical Study on LLM-based Green Code Generation}
This study~\cite{vartziotis2024learn} examines whether generative AI models can generate code that meets sustainability objectives, focusing on metrics like runtime, memory usage, energy consumption, and floating-point operations (FLOPs). Results show that AI-generated code does improve sustainability metrics upon optimisation requests but typically falls short of human-coded solutions from LeetCode. Our work uses the same metrics but instead we report the metrics directly whereas Vartziotis et al. merge them to 'Green capacity'. Also, we use DeepSeek-Coder and Code Llama instead of GitHub Copilot~\cite{dohmke2022github}, Amazon CodeWhisperer~\cite{amazon2022}, and OpenAI ChatGPT-3~\cite{openai_introducing_2022}.

\subsection{Energy Efficiency of the Code Generated by Code Llama}
This work~\cite{cursaru2024controlled} is similar to our setup and measures the energy efficiency of code generated by Code Llama. The paper shows that the performance is highly dependant on the coding language and problem. Generally speaking, human-generated code is more energy-efficient than Code Llama's output, even when prompted for energy-efficient code. Our work differs as we do not examine the regular output but rather prompt for green coding principles, next to that we extend this work by including the model DeepSeek-Coder. 

\section{Methodology} \label{sec:methodology}
Apart from energy consumption, to investigate the performance of the code generated by LLMs, we use the same metrics as Vartziotis et al. (2024)~\cite{vartziotis2024learn}: runtime, peak memory usage, and the number of floating-point operations (flops). Though we report all these metrics for compatibility with related work, we focus on energy consumption to answer our research questions. We also describe the selection of LLMs, code prompts, and best practices to test prompt modification.

We use Perf and GNU's Not Unix (GNU) \texttt{time} command to measure energy consumption and resource management.

\textbf{Perf} is not primarily an energy profiler, but it offers energy profiling functionalities. Perf supports a list of measurable events, that vary with each processor type and model \cite{perfmanual}. Sources of the events are kernel counters, the processor, and the processor's Performance Monitoring Unit (PMU). Perf is capable of measuring energy-related events. On Intel-based systems, data of the energy events are provided with the interface Running Average Power Limit (RAPL)~\cite{RAPL}.

We use Perf because it works independently of programming languages as it measures the statistics per single command. Perf allows repeated measurements of the same event and captures the event's output together with the elapsed time. We use the event \texttt{power/energy-pkg/} to gather the energy consumption of a command. This Perf event captures the energy consumption, runtime and \textbf{flops} of the whole CPU package and returns the \textbf{energy data} in Joules~\cite{hahnel2012measuring}.

To measure the \textbf{peak memory usage} of each command, we use the GNU tool \texttt{time}. This tool runs a command and returns a summary of used system resources. This includes the peak memory usage of the command. We use time because it measures a command's peak memory consumption and is programming language independent.

\subsection{Selecting Large Language Models}
Of the LLMs used by Vartziotis et al.~\cite{vartziotis2024learn}, only ChatGPT-3 is supported with an API at the time of this writing. ChatGPT-3 can be integrated via a paid API and is therefore unavailable for this study. Amazon CodeWhisperer and Github Copilot are both an extension of an integrated development environment.\\
This study has access to public LLMs, a list of available LLMs can be found on GitHub\footnote{\url{https://github.com/eugeneyan/open-llms}, accessed on 27-03-2024.}.
We select DeepSeek-Coder-Base and DeepSeek-Coder-Instruct due to their strong performance. Because of its high performance in previous work~\cite{guo2024deepseek} we choose the 33B parameter model. The Code Llama-base models rank second-best~\cite{guo2024deepseek}. While only the 33B parameter version was considered in these results, there is evidence that the 70B model outperforms the 33B model in both the HumanEval and MBPP benchmarks~\cite{roziere2023code}. Thus, we use the 70B versions of the Code Llama model: Code Llama 70B, Code Llama - Instruct 70B, and Code Llama - Python 70B.

\begin{table}[h!]
\centering
\begin{tabular}{lcc}
\toprule
\textbf{Model} & \textbf{Type} & \textbf{Parameter Size} \\ 
\midrule
DeepSeek-Coder-Base & Base & 33B \\ 
DeepSeek-Coder-Instruct & Instruction-tuned & 33B \\ 
Code Llama & Base & 70B \\ 
Code Llama - Instruct & Instruction-tuned & 70B \\ 
Code Llama - Python & Python-tuned & 70B \\ 
\bottomrule
\end{tabular}
\caption{Overview of Selected Large Language Models}
\label{tab:llm_overview}
\end{table}

\subsection{Selecting code problems}
LeetCode\footnote{\url{https://leetcode.com/problemset/}, accessed on 02-03-2024.} is a platform that provides code problems, categorised by difficulty levels: easy, medium, and hard. We choose this platform for compatibility with related work~\cite{vartziotis2024learn}. Each problem is provided with a starting code snippet stating, if applicable for the programming language, the class function name and inputs of the function. To create a code prompt we copy the code problem text description and corresponding starting code snippet. We chose Python as the programming language for the code problems, due to its usage in a wide variety of domains and popularity~\cite{srinath2017Python}. \\
We select one problem for each difficulty level, whereas Vartziotis et al~\cite{vartziotis2024learn} pick 6 problem sets. Each problem is provided with constraints, so we have tested the solutions with test cases that are as large as possible while still adhering to the constraints. We test the code problems with large test cases to increase the runtime and with that, we increased the granularity, since short energy measurements can be inaccurate~\cite{hahnel2012measuring}.Table~\ref{tab:overview_code_problems} shows an overview of the code problems.
\renewcommand{\tabcolsep}{1.1pt}

\begin{table}[htb]
    \centering
    \begin{tabular}{c|c|c|c}
        \toprule
        \textbf{Level} & \textbf{Easy} & \textbf{Medium} & \textbf{Hard} \\ 
        \midrule
        Code Problem & 
        \href{https://leetcode.com/problems/assign-cookies/description/}{Assign Cookies} & 
        \href{https://leetcode.com/problems/sort-list/}{Sort List} & 
        \href{https://leetcode.com/problems/median-of-two-sorted-arrays/description/}{Median of Two Sorted Arrays} \\ 
        \bottomrule
    \end{tabular}
    \caption{Overview of the different code problems}
    \label{tab:overview_code_problems}
    \vspace{-6mm}
\end{table}

\subsection{Prompt modification}
Previous work~\cite{vartziotis2024learn} modifies prompts to generate code that consumes less energy. Their method involves adding a sentence at the beginning of the code prompt, specifying that the solution must be energy-optimised. A LLM can provide an optimised solution if it understands how to enhance the code to consume less energy. This assumes that the LLM understands the problem and knows how to improve the outcome. From a user perspective, this approach is straightforward and does not require any knowledge of code optimisation. \\
We use this method as a baseline for testing prompt modifications. In addition to this approach, we examine best practices for writing energy-efficient Python code. \cite{koedijk2022finding} examines programs that solve the same problem with varying energy consumption using the same programming language. The study identifies techniques for writing more energy-efficient code. The study shows that in Python, the use of for-loops is more efficient than while-loops. The study also shows that using libraries instead of writing the code yourself reduces energy consumption. We incorporate these findings into our prompt modifications. Each finding is tested separately by adding a sentence to the beginning of the code prompt. The sentences together with their labels are shown in Table~\ref{fig:optimisation_prompts}.  

\begin{table}[H]
\centering
\begin{tabular}{ll}
\toprule
\textbf{Category} & \textbf{Description} \\ 
\midrule
Energy & Give me an energy-optimised solution for this problem \\ 
Libraries & Use library functions in the following problem \\ 
For loop & Use a for-loop instead of a while-loop in the following problem \\ 
\bottomrule
\end{tabular}
\caption{Overview of the prompt optimisation sentence types.}
\label{fig:optimisation_prompts}
\end{table}

\vspace{-8mm}
\subsection{Code similarities}\label{section:code_similarities}
We examine the similarity between the generated code for non-optimised and optimised code prompts addressing the same problem using \texttt{pycode-similar}, a plagiarism detector for Python code. The tool compares the abstract syntax tree (AST) of the different code solutions~\cite{pycode_similar}, by normalising the AST to remove unhelpful attributes, such as print statements, and then using \texttt{difflib} to determine the differences between the ASTs. 

\section{Experimental setup and results} \label{sec:experiments}
The LLMs are hosted on an Oracle server, as the models require significant hardware resources. We access the Oracle server via a command line interface and transfer the output to our local machine. The measurements are conducted on a laptop with an 11th Gen Intel Core i5-1135G7, Ubuntu 22.04.4 LTS and kernel version 6.5.0-35-generic. Nota bene, because we analyse the energy consumption of the code generated by the code LLMs, not the models themselves, we do not use a GPU in our experimental setup. The Python programs used in our study are available \href{https://github.com/tcappendijk/Greencode_framework}{on GitHub}. 

To minimise background tasks, we use a terminal and prevent the system from updating by putting the device in 'airplane mode' during the experiments. Energy consumption and runtime are measured 50 times for each prompt. The peak memory usage measurement is repeated 50 times to account for variations caused by cache effects and cold starts~\cite{pereira2017energy}. The energy consumed by main memory depends on the access pattern of the program. As total memory consumption does not consider this, we use peak memory in alignment with related work~\cite{vartziotis2024learn}. The total number of floating point operations is not influenced by the Python process, even when floating point operations are executed, thus we test and verify this once.

\noindent
\textit{Output Modification.} \label{experiments:output_modification}
This is a manual step required because none of the LLM code outputs can be used directly due to one or more of the following issues, which results in incorrect Python code: repetition of the code prompt in the output, usage of libraries without importing them, irrelevant code after the code solution, and omission of the initial code snippet provided in the code prompt. Here, the "code snippet" refers to the starting point of the prompt, which serves as the basis for the LLM's continuation.

We solely adjust the LLMs' code output when it does not influence the scope of the class or method relevant to the code problem. This ensures that modifications do not affect the solution itself. LLMs are probabilistic models, so we examine their outputs across multiple runs and find no difference in code output across different runs (see Git).

\section{Results}
Each configuration consists of one LLM together with one prompt. The prompts are derived from three different code problems and four different prompt types (default $+$ three optimisation sentences), resulting in a total of sixty configurations (twelve prompts $\times$ five LLMs). We apply the Mann-Whitney U test and compare configurations without optimisation against those with optimisation for the same code problem and LLM, as seen in various previous work~\cite{koedijk2022finding, cursaru2024controlled}. This comparison is based on differences in energy consumption, runtime, the total number of floating-point operations, and peak memory usage.

We use the Mann-Whitney U test because it is independent of the underlying distributions of our data, assuming that these distributions are similarly shaped~\cite{TQMP4-1-13}. The null hypothesis of the Mann-Whitney U test is that energy dataset $x$ belongs to the same energy population as energy dataset $y$~\cite{TQMP4-1-13}. The alternative hypothesis is determined by the variant of the Mann-Whitney U test. The two-sided Mann-Whitney U test has an alternative hypothesis that energy dataset $x$ does not belong to the same population as energy dataset $y$. The alternative hypothesis of the one-sided Mann Whitney is that energy dataset $x$ is stochastically larger than energy dataset $y$. We use the one-sided variant twice because we wish to say whether energy dataset $x$ is stochastically larger or smaller than energy dataset $y$. The first application has the null hypothesis that energy dataset $x$ belongs to the same energy population as energy dataset $y$. The alternative hypothesis is that energy dataset $x$ is stochastically larger than energy dataset $y$. We test this with an $alpha$-value of 0.01 because we wish to reduce the chance of falsely rejecting the null hypothesis. If we cannot reject the null hypothesis, the $p$-value is greater than the $alpha$-value, we cannot say that energy dataset $x$ is stochastically greater than energy dataset $y$ and test the opposite. We use the null hypothesis that energy dataset $y$ is from the same population as energy dataset $x$. The alternative hypothesis is that energy dataset $y$ is stochastically larger than energy dataset $x$. If the null hypothesis cannot be rejected in the second test either, we deem the results inconclusive.

\begin{figure}[htp]
  \hspace*{-2cm}
    \centering
    \includegraphics[width=1.0\columnwidth]{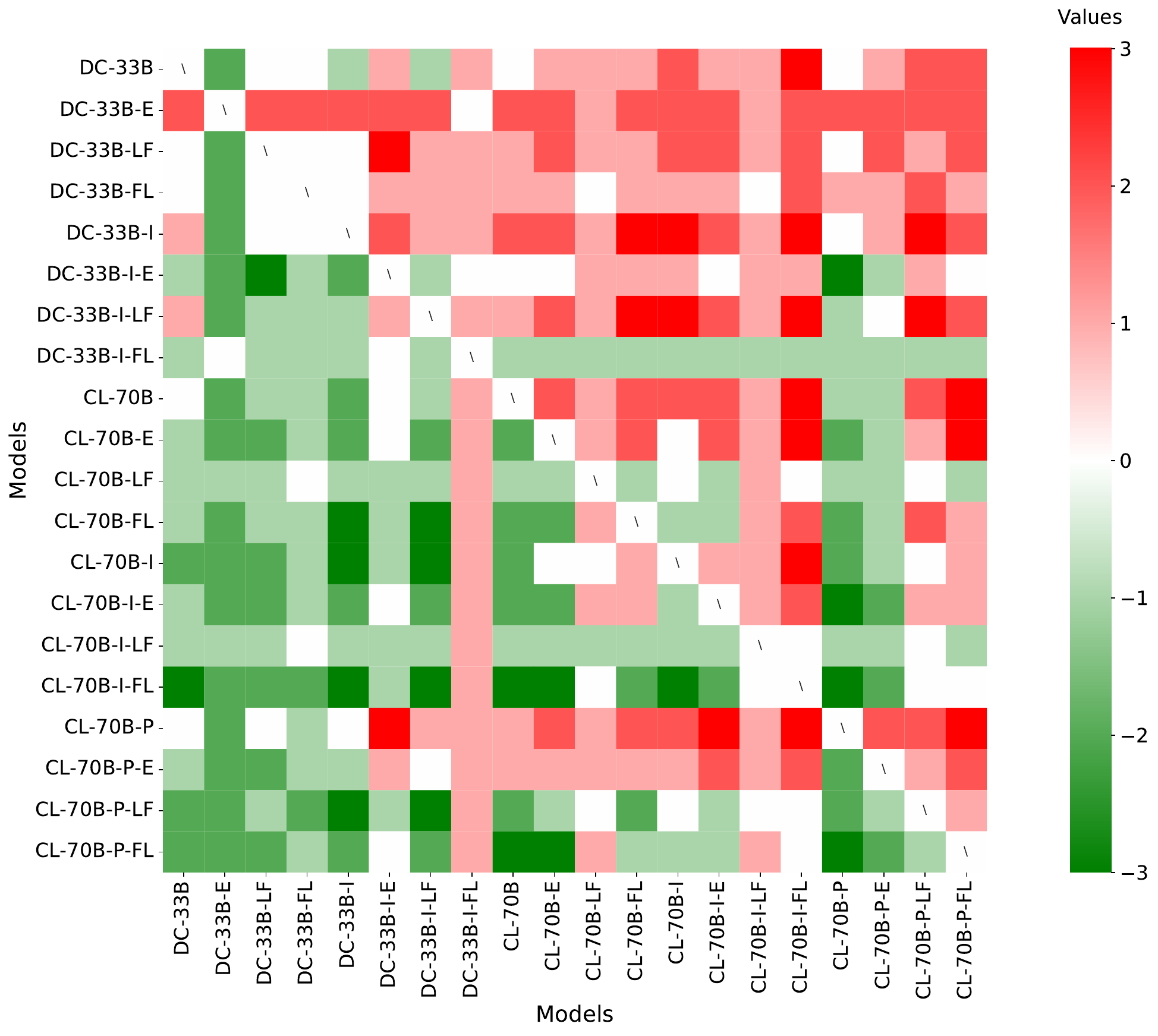}
    \caption{Comparison of the different LLMs and their prompts for all the problems. We abbreviate DeepSeek Coder to DC and Code Llama to CL. B represents the Base prompt, E represents the prompt for Energy efficiency, LF the prompt for Library Functions and finally FL the use of For Loops. Green shows that the configuration on the row consumes less energy than the configuration on the column for all the problems, red, the opposite.}
\label{fig:mann_whitney_u_test_known_values_all_three_problems}
%\vspace{-15mm}
\end{figure}

\begin{figure}[htb]
    \hspace*{-0.5cm}
    \centering
    \includegraphics[width=1.0\columnwidth]{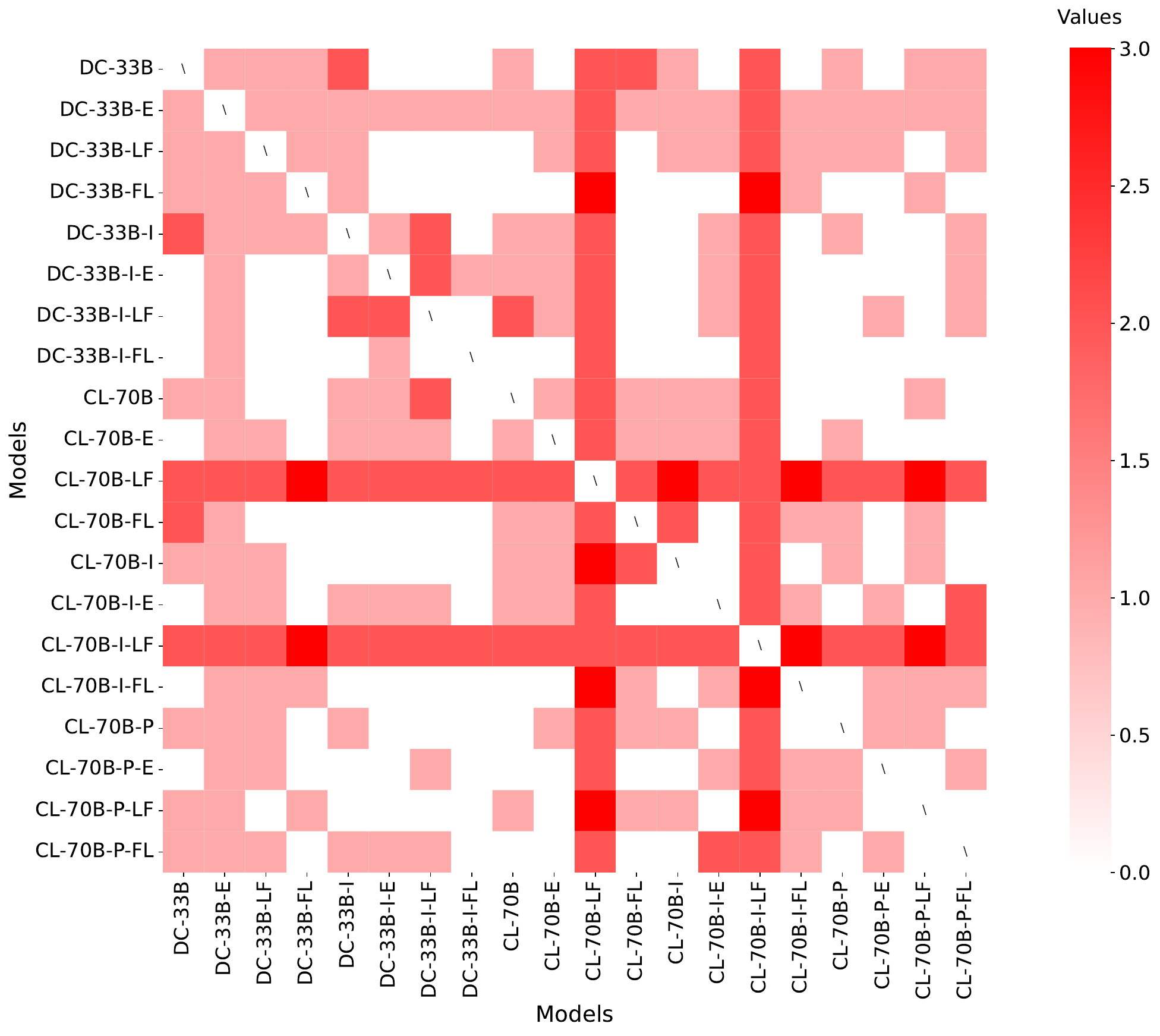}
    \caption{Comparison of the different LLMs and their (optimised) prompt for all the problems. We abbreviate DeepSeek Coder to DC and Code Llama to CL. B represents the Base prompt, E represents the prompt for Energy efficiency, LF the prompt for Library Functions and finally FL the use of For Loops. The values indicate the number of times the unknown value occurred between the two configurations.}
    \label{fig:mann_whithney_u_unknown_values}
\end{figure}

We calculate a score for every combination to get an overview of the energy consumption of the different configurations across the code problems. A value of one is assigned if the configuration in the row consumes more energy than the configuration in the column and a value of minus one is assigned if the opposite holds. No value is assigned otherwise, given we have three code problems the numbers may add up to -3 or 3. Figure~\ref{fig:mann_whitney_u_test_known_values_all_three_problems} shows the results, and Figure~\ref{fig:mann_whithney_u_unknown_values} shows occurrence of unknown comparison results.

\subsubsection*{Base Prompt Versus Optimised Prompt}
We assess the difference between the base- and optimised prompt. The metrics used are similarity, energy, memory, flops, and runtime. We are not interested in the value itself, but rather in the difference between the base- and optimised prompt. The similarity tool computes this difference at code level. For the other metrics, we report the difference as a percentage of the base value:
\setlength{\abovedisplayskip}{3pt}
\setlength{\belowdisplayskip}{3pt}
\begin{equation}
    \frac{\texttt{optimised} - \texttt{base}}{\texttt{base}} \times 100
    \label{eq:performance-improvement}
\end{equation}
The total number of floating point operations is measured once so we use the result directly. The result for the metrics energy, runtime, and peak memory consumption is a set of fifty data points. To calculate the percentage difference, we use the mean of each dataset.

\begin{table}[htb]
\centering
\begin{tabular}{lcccccc}
\toprule
 & \multicolumn{2}{c}{Sort List} & \multicolumn{2}{c}{Assign Cookies} & \multicolumn{2}{c}{\begin{minipage}{.25\columnwidth}\centering Median of Two Sorted Arrays\end{minipage}} \\
\cmidrule(lr){2-3} \cmidrule(lr){4-5} \cmidrule(lr){6-7}
 & LF & FL & LF & FL & LF & FL \\
\midrule
Code Llama-70b & 0 & 1 & 1* & 0 & 0 & 0 \\
Code Llama-70b-Instruct & 0 & 1 & 1* & 0 & 0 & 0 \\
Code Llama-70b-Python & 1 & 0 & 1* & 0 & 1* & 0 \\
deepseek-coder-33b-base & 0 & 0 & 1* & 0 & 1* & 0 \\
deepseek-coder-33b-instruct & 0 & 0 & 1* & 0 & 1* & 0 \\
\bottomrule
\end{tabular}
\caption{Overview that shows whether an LLM actually applied the optimisation prompts for \texttt{library functions} -- LF, and \texttt{for-loop} -- FL, respectively for each code problem. A 0 indicates that the optimisation is not present in the code solution and a 1 is the opposite. * means that the same optimisation code was present in both the optimised solution and the base solution.}
\label{tab:difference_base_optimisation_lf_fl}
\end{table}

\begin{table*}[ht!]
    \centering
    \begin{adjustbox}{max width=\textwidth}
    \begin{tabular}{@{}l|l|ccccc|ccccc|ccccc@{}}
    \toprule
     Model & Prompt & \multicolumn{5}{c}{\textbf{Assign Cookies}} & \multicolumn{5}{c}{\textbf{Sort List}} & \multicolumn{5}{c}{\textbf{Median of Two Sorted Arrays}} \\
    \cmidrule(lr){3-7} \cmidrule(lr){8-12} \cmidrule(lr){13-17}
     &  & Sim. & Energy & Mem. & FLOPs & Runtime & Sim. & Energy & Mem. & FLOPs & Runtime & Sim. & Energy & Mem. & FLOPs & Runtime \\
    \midrule
    & energy & 90.6 & 13.4 & -0.1 & 0.0 & 10.6 & 66.9 & 1.7 & 0.0 & 0.0 & 0.9 & / & / & / & / & / \\
    DC-33b-base & lib func & 90.6 & 8.8 & 0.1 & 0.0 & 11.0 & 98.6 & -0.3 & 0.0 & 0.0 & 0.1 & 100.0 & -3.1 & 0.0 & 0.0 & 0.0 \\
    & for-loop & 100.0 & -3.8 & 0.1 & 0.0 & 0.4 & 74.1 & -0.2 & -0.1 & 0.0 & -1.5 & 91.2 & 465.5 & 1.0 & 0.0 & 496.6 \\ \hline
    & energy & 100.0 & 0.2 & 0.1 & 0.0 & -0.4 & 92.2 & -0.8 & 0.0 & 0.0 & 0.8 & 45.7 & -17.3 & 1.3 & 0.0 & -19.1 \\
    DC-33b-inst & lib func & 100.0 & 0.3 & 0.0 & 0.0 & -0.3 & 100.0 & -1.3 & -0.3 & 0.0 & -0.6 & 100.0 & 0.5 & 0.0 & 0.0 & -0.1 \\
    & for-loop & 64.5 & -26.8 & -8.3 & 0.0 & -26.9 & 93.0 & 2.9 & -0.2 & 0.0 & 3.5 & 45.7 & -17.4 & 1.1 & 0.0 & -19.3 \\ \hline
    & energy & 96.9 & 4.0 & -0.1 & 0.0 & 0.2 & 92.9 & -2.0 & 0.1 & 0.0 & -1.2 & 88.2 & -3.3 & -0.5 & 0.0 & -0.3 \\
    CL-70b-inst & lib func & 96.9 & -2.1 & 0.2 & 0.0 & 0.4 & 97.3 & /   & / & / & /   & / & / & / & / & /   \\
    & for-loop & 96.9 & -1.7 & 0.0 & 0.0 & 0.3 & 77.7 & -21.6 & 24.1 & 0.0 & -19.1 & 88.2 & -1.8 & -0.3 & 0.0 & 0.0 \\ \hline
    & energy & 96.9 & -0.5 & -0.2 & 0.0 & -0.2 & 92.9 & -0.9 & 0.3 & 0.0 & -1.2 & 88.2 & -3.2 & 0.2 & 0.0 & 0.0 \\
    CL-70b & lib func & 96.9 & -6.1 & -0.2 & 0.0 & -0.4 & 97.3 & /   & / & / & /   & / & /   & / & / & /   \\
    & for-loop & 96.9 & -4.2 & -0.2 & 0.0 & 1.0 & 77.7 & -21.7 & 24.2 & 0.0 & -19.3 & 88.2 & -2.3   & 0.1 & 0.0 & 0.0   \\ \hline
    & energy & 100.0 & 0.3 & 0.0 & 0.0 & -0.1 & 85.2 & -1.2 & 0.0 & 0.0 & -2.4 & 100.0 & -4.4 & -0.4 & 0.0 & 0.0 \\
    CL-70b-Py & lib func & 79.4 & -9.7 & 0.0 & 0.0 & -9.6 & 77.4 & -59.4 & -0.1 & 0.0 & -60.0 & 100.0 & 0.3 & -0.1 & 0.0 & 0.0 \\
    & for-loop & 82.3 & -5.0 & 0.0 & 0.0 & -9.6 & 76.5 & -59.7 & 0.0 & 0.0 & -60.1 & 100.0 & -4.9 & -0.2 & 0.0 & -0.1 \\
    \bottomrule
    \end{tabular}
    \end{adjustbox}
    \label{tab:base_code_prompt_against_optimised_code_prompts_all_problems}
        \caption{The base code prompt solution compared to the optimised prompt solutions across coding problems. Negative percentages indicate that the base code prompt solution has a higher value and vice-versa for positive values. The mean value of 50 samples is taken for the metrics energy, peak memory consumption, and runtime. We write a missing value "/" if the model does not output for the prompt. Model abbreviations: DC = deepseek-coder-33b, CL = CodeLlama-70b.}
    % \caption{Metrics for different models across coding problems. Missing values are indicated with "/". Model abbreviations: DSC = deepseek-coder-33b, CL = CodeLlama-70b.}
    % \label{tab:metrics_models}
\end{table*}

\begin{figure}[h]
    \hspace*{-2cm}
    \centering
    \includegraphics[width=1.0\columnwidth]{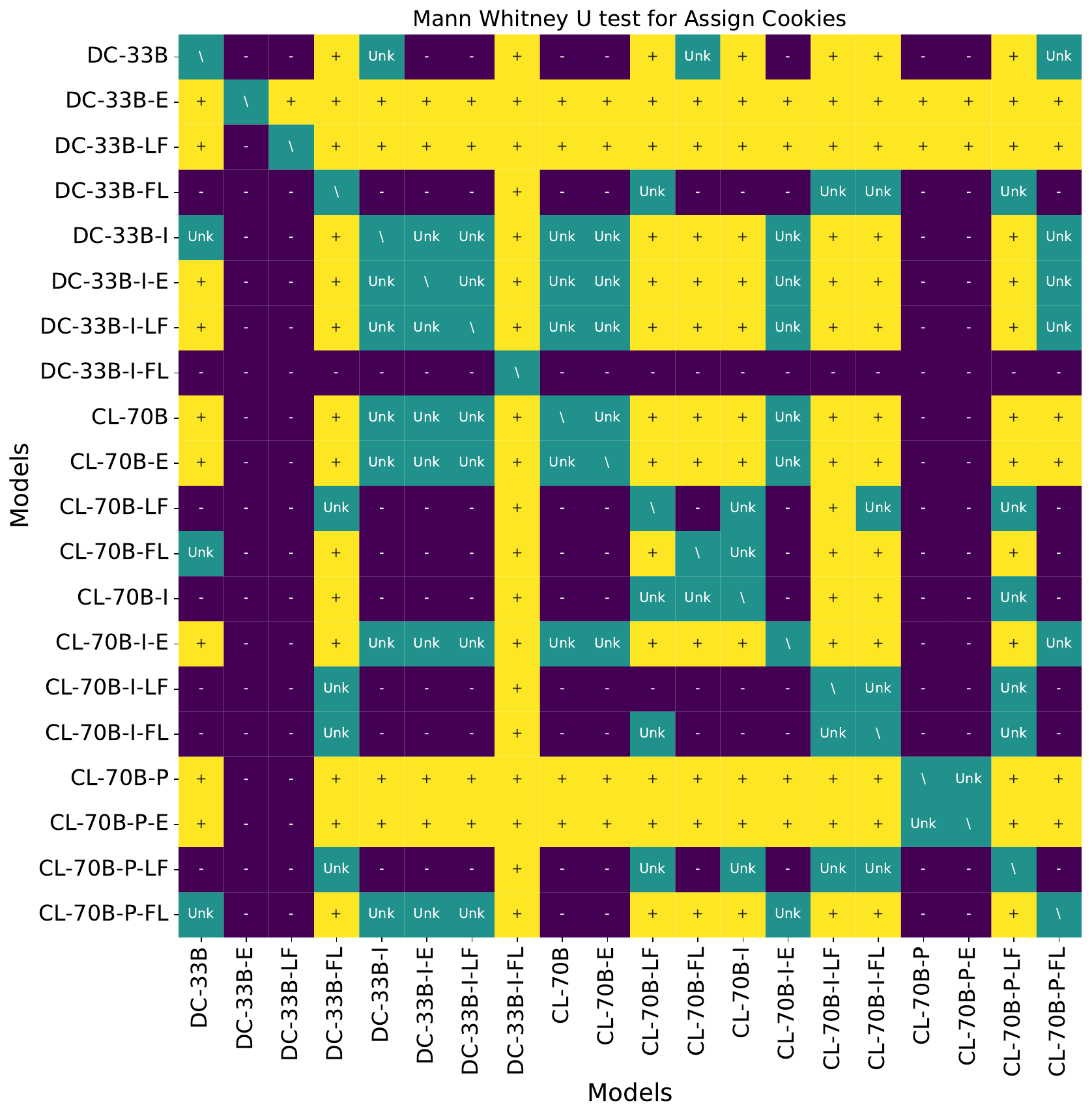}
    \caption{Comparison of the LLMs and their (optimised) prompt for the Assign Cookies problem. A + means that the configuration on the row consumes more energy than the configuration on the column, a - is the opposite and Unk stands for Unknown meaning that both null hypotheses cannot be rejected. Model abbreviations: DC = deepseek-coder-33b, CL = CodeLlama-70b, I = instruction-tuned, Py = Python-version, E = prompt for energy-effiency, LF = library function \& FL = For-Loop.}
    \label{fig:mann_whitney_u_test_assign_cookies}
\end{figure} 

\begin{figure}[h]
    \hspace*{-1cm}
    \centering
    \includegraphics[width=1.0\columnwidth]{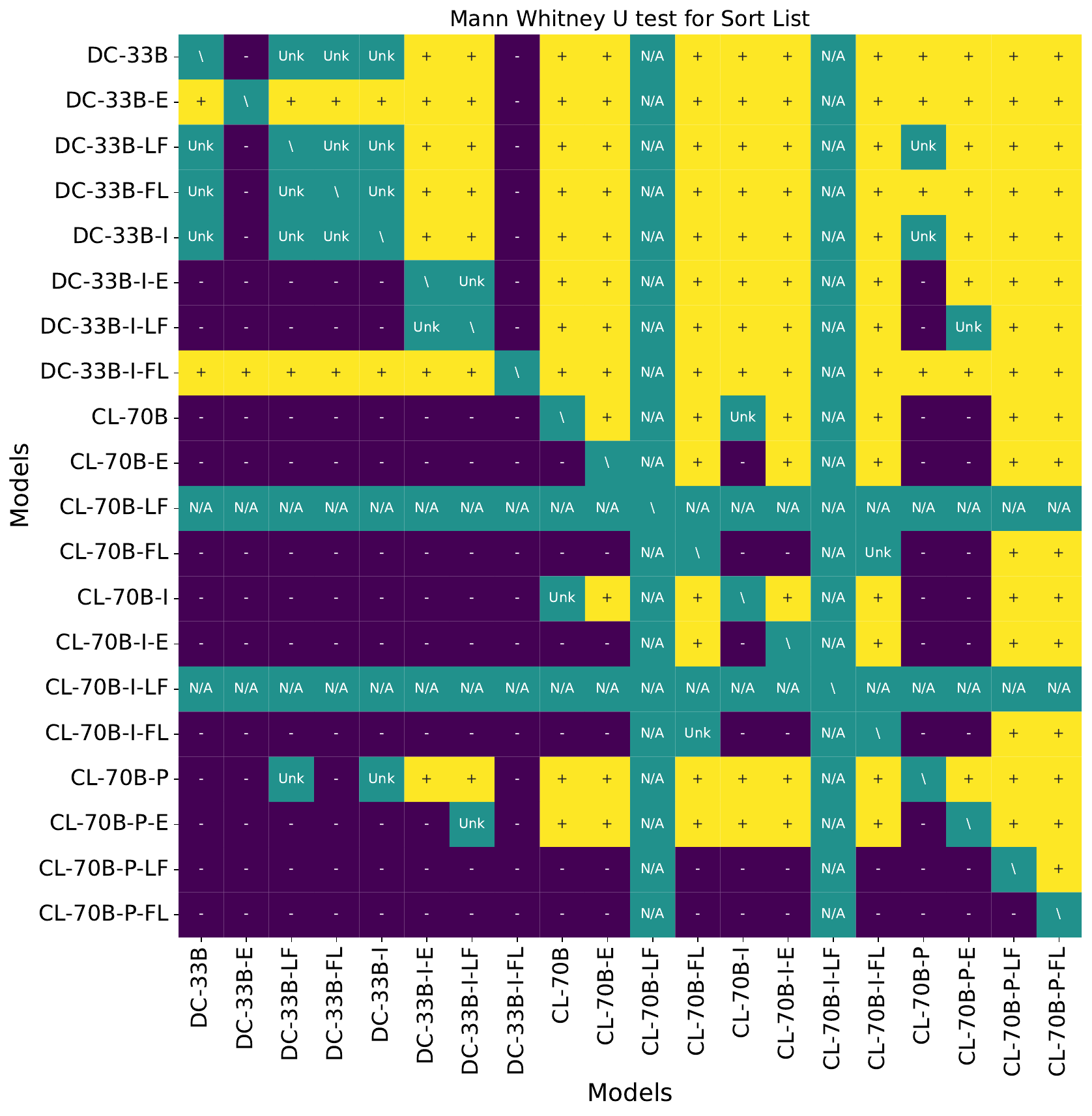}
    \caption{Comparison of the LLMs and their (optimised) prompt for the Sort List problem. 
    A + means that the configuration on the row consumes more energy than the configuration on the column, a - is the opposite and Unk stands for Unknown meaning that both null hypotheses cannot be rejected. Model abbreviations: DC = deepseek-coder-33b, CL = CodeLlama-70b, I = instruction-tuned, Py = Python-version, E = prompt for energy-effiency, LF = library function \& FL = For-Loop.}
    \label{fig:mann_whitney_u_test_sort_list}
\end{figure}

\begin{figure}[h]
    \hspace*{-2cm}
    \centering
    \includegraphics[width=1.0\columnwidth]{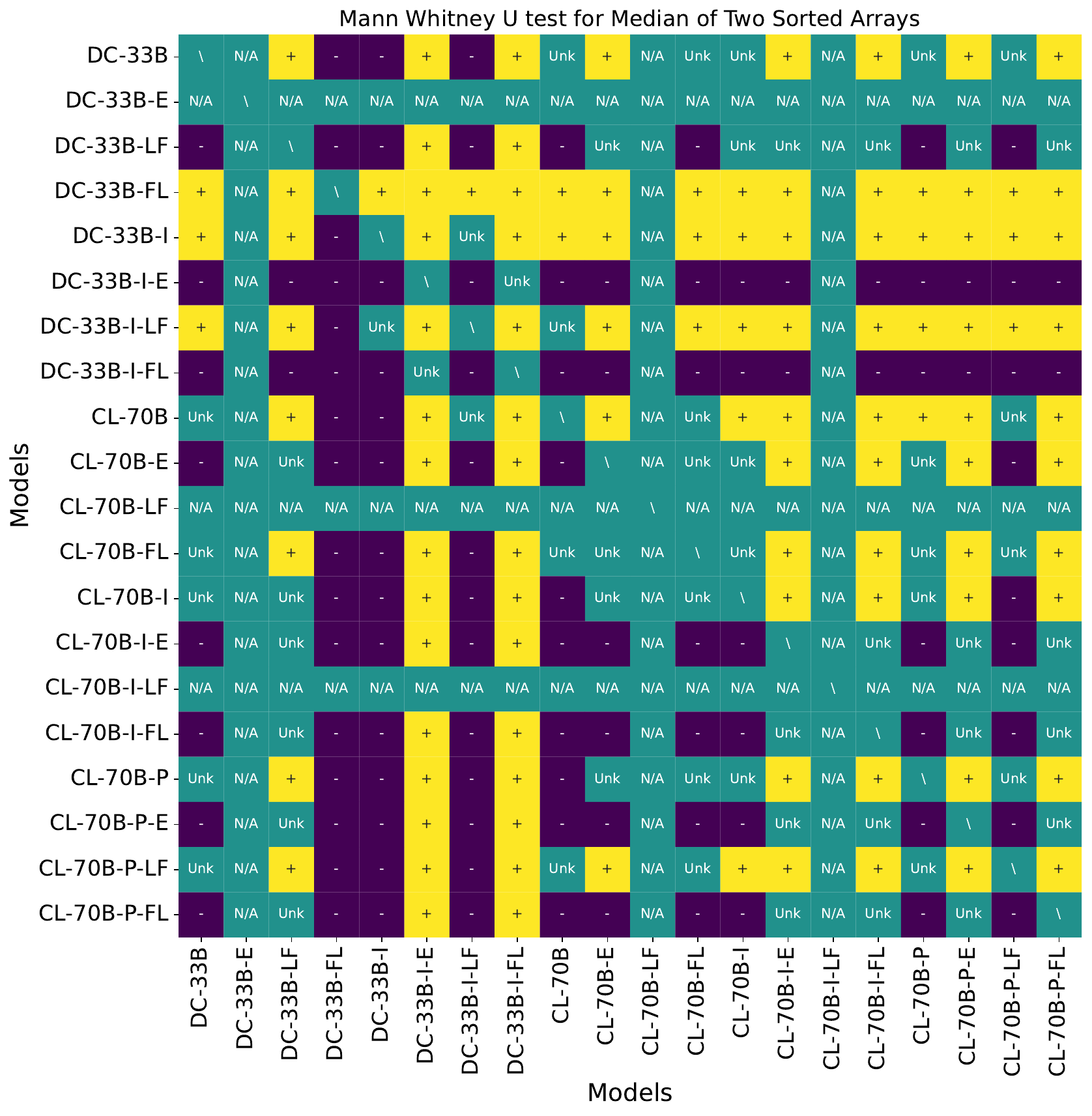}
    \caption{Comparison of the LLMs and their (optimised) prompt for the Median of Two Sorted Arrays. 
    A + means that the configuration on the row consumes more energy than the configuration on the column, a - is the opposite and Unk stands for Unknown meaning that both null hypotheses cannot be rejected. Model abbreviations: DC = deepseek-coder-33b, CL = CodeLlama-70b, I = instruction-tuned, Py = Python-version, E = prompt for energy-effiency, LF = library function \& FL = For-Loop.}
    \label{fig:mann_whitney_u_test_median_of_two_sorted_arrays}
\end{figure}

The \texttt{library functions} and \texttt{for-loop} optimisation prompts give an exact instruction on what the LLM must apply to generate code with lower energy consumption. We looked at the code generated with these optimisation prompts and made an overview of the configurations in which the optimisation is applied. This overview is shown in Table~\ref{tab:difference_base_optimisation_lf_fl}. The value 1 indicates that the optimisation is present in the code solution. This means that a minimum of one instance is present in the solution. The * indicates that the optimisation is present in the base solution, this is the code prompt without an added optimisation sentence. The value 0 indicates that the optimisation is not present in the code output.

\section{Discussion} \label{sec:discussion}
When analysing the results of prompt modification, we find something unexpected. Table~\ref{tab:base_code_prompt_against_optimised_code_prompts_all_problems} shows that the Code Llama-70b-Python model has no difference in code output for the Median of Two Sorted Arrays code problem as the similarity is 100. There is however a difference in the mean value of the energy datasets. This difference is 4.4\%, 0.3\% and 4.9\% between the base prompt and the \texttt{energy}, \texttt{library functions}, and \texttt{for-loop} optimisation prompt respectively. As the code output is the same for all prompts, we think this difference in energy consumption is caused by the noise of background processes, which we discuss in our limitations.\\
Comparison for the same model and code problem in Figure~\ref{fig:mann_whitney_u_test_median_of_two_sorted_arrays} we find that the Mann Whitney U test assigns a plus (+) to the base prompt compared with the \texttt{energy} or the \texttt{for-loop} optimisation. The correct value for this comparison is unknown (Unk) because there should not be a significant difference in energy consumption between two similar code solutions. This example illustrates that we cannot draw conclusions with the Mann Whitney U test about the energy consumption difference of two solutions. 

\subsection{Energy Consumption}\label{discussion:energy_consumption}
\noindent
When comparing the mean value of the optimisation prompt energy dataset against the base prompt energy dataset, we see some notable differences. We look at percentage differences larger than 15\% and try to find their cause. The code solution corresponding to the base prompt and the optimisations prompts for all LLMs are available in our Git repository\footnote{\url{https://github.com/tcappendijk/Greencode_framework}}.

We start with the results for the Sort list code problem, collected in Table~\ref{tab:base_code_prompt_against_optimised_code_prompts_all_problems}. We find that the \texttt{for-loop} optimisation prompt consumes 21.7\%, 21.6\%, and respectively 59.7\% less energy compared to the base prompt for the LLMs Code Llama-70b, Code Llama-70b-Instruct, and Code Llama-70b-Python. 
The optimised solutions convert the linked list to an array and use a library function to sort the array. The resulting sorted array must be converted back to a linked list. For this Code Llama-70b and Code Llama-70b-Instruct use a \texttt{for-loop}, the Code Llama-70b-Python uses a \texttt{while-loop}. The computationally intensive aspect of the solution is list sorting. Reconstructing the list into a linked list via a loop contributes to a minor portion of execution time and thus has small impact on the energy consumption compared to sorting using a library function or Python code. Figure~\ref{fig:mann_whitney_u_test_sort_list} shows that the base prompt for Code Llama-70B-Python uses more energy than the base prompts for Code Llama-70B and Code Llama-70B-Instruct. Next to that, the for-loop optimised solution of Code Llama-70B-Python uses less energy than the for-loop-optimised solutions for Code Llama-70B and Code Llama-70B-Instruct. This means the for-loop optimisation for Code Llama-70B-Python has the largest decrease in energy consumption, as its base prompt uses the most energy, but its optimised prompt uses the least.

The \texttt{library functions} prompt optimisation consumes 59.4\% less energy compared to the base prompt for the Code Llama-70b-Python model. 
We find that the difference on a code level between the base on the optimisation is the same as for the Code Llama-70b-Python LLM with the \texttt{for-loop optimisation}. The two DeepSeek-Coder LLMs do not improve. Their base code solution sorts the linked list in Python code, the optimised solutions do not change to sorting with a library function.

For the Assign cookies code problem, we find that the DeepSeek-coder-33b-instruct LLM with the \texttt{for-loop} prompt optimisation consumes 26.8\% less energy compared to the base prompt. On a code level we find that the base version updates a counter, and indexes with this counter in both input lists. The \texttt{for-loop} optimised solution, does not loop through the sorted input lists. Instead, it uses the -1 index notation to get the last element of the list. The break condition of the loop is when both input lists are empty. This optimised solution does not contain a for-loop. Instead, the difference between the base and the optimised code solution is within using the library function \texttt{pop()} and index method.

For the Median of Two Sorted Arrays code problem, we look at Table~\ref{tab:base_code_prompt_against_optimised_code_prompts_all_problems}. For the DeepSeek-Coder-33b-base LLM, we find that the \texttt{for-loop} optimisation has an increase in energy consumption of 465.5\% compared to the base prompt. 
We find that the base solution merges the two input lists with a + operation and sorts the result with a library function. The optimised solution is merging and sorting the two input arrays in native Python code and uses three while-loops for this. The DeepSeek-Coder-33b-instruct LLM compared with the base prompt shows a decrease of 17.3\% and 17.4\% in energy consumption for the \texttt{energy} and \texttt{for-loop} optimisation prompts respectively. 
The code solutions for the \texttt{energy} and \texttt{for-loop} optimisation prompts are the same. 
There is a difference in the code output between the base and the optimised code solutions. We find that the base solution merges the two input lists. This input list is sorted and based on the length of this list, the median value is retrieved. For the \texttt{for-loop} and \texttt{energy} optimisation, the input arrays are not merged and sorted. This solution searches the middle point of the two input arrays, as if they were combined, and determines the median based on the values around this center. This code solution does not contain a \texttt{for-loop}, instead containing a \texttt{while-loop}.

\subsection{Total Number Of Floating Points}
We measure the percentage difference in the total number of floating point operations between the base solution and an optimised solution. Across all three code problems, this percentage difference is zero. This means that we cannot use the total number of floating point operations to asses a difference between configurations.

\subsection{Peak Memory Consumption}
The difference in peak memory consumption between the base solution and the optimised solution is between -0.5\% and 1.3\% with two exceptions. These exceptions are 24.1\% and 24.2\%, indicating that the base prompt consumes less peak memory than the optimised prompt. The configurations that achieved these exception values are the LLMs Code Llama-70b-Instruct and Code Llama-70b with the \texttt{for-loop} optimisation prompt for the Sort List code problem. Because of the minimal difference, we cannot use peak memory consumption to asses a difference between configurations.

\subsection{Answering the Research Questions}
%\todo[inline]{TODO Ana: please proofread the text below until "LLM Code Output Similarity"; Ana: Done! I've left a couple of TODOs}
Regarding RQ~\ref{question:2.1}, we find that the \texttt{for-loop} optimisation prompt most frequently triggers the LLMs to generate solutions that consume less energy compared to the base solution. These solutions do not have to include a \texttt{for-loop}, they may perform better due to the use of library functions. However, the \texttt{for-loop} optimisation does not consistently show a decrease in energy consumption across all LLMs or for the same LLM across all code problems. There is a case where the \texttt{for-loop} optimisation led to a 465.5\% increase in energy consumption. We find that the base solution merges the two input lists with a + operation and sorts the result with a library function. The optimised solution merges and sorts the two input arrays in native Python code and uses three while loops. Therefore, we cannot conclude whether an optimisation prompt would consistently decrease energy consumption in all LLMs and code problems. This means we cannot determine with 100\% confidence which prompt optimisation consumes less energy for a given LLM. 

Regarding RQ~\ref{question:2.2}, the Sort List code problem most frequently shows a decrease in energy consumption of more than 15\% with prompt optimisation. However, this code problem does not show this decrease in all three optimisation prompts. 

The answer to RQ~\ref{question:2.1} is that a specific combination of prompt optimisation, LLM and code problems can lead to a decrease in energy consumption. However, no single optimisation prompt consistently decreases the energy consumption for the same LLM across the different code problems. We have a similar answer for RQ~\ref{question:2.2}. Based on the experimental results, we cannot determine the relation between coding problems and prompt optimisation for energy consumption.

\subsection{LLM Code Output Similarity}
We find that the similarity is 100\% across all runs, with one exception. We investigated by checking the parameters of the LLMs. The Deepseek-coder LLMs are implemented via \href{https://huggingface.co/}{HuggingFace}, which has a \texttt{do\_sample} parameter which is by default false, we did not adjust this value, meaning that the LLM uses greedy decoding~\cite{HuggingfaceGeneration}. This decoding strategy picks the token with the highest probability as the next token and the result is a deterministic output~\cite{HuggingfaceTextGenerationStrategies}. The Code Llama LLMs use a parameter called \texttt{temperature}, which regulates the degree of randomness in the output of a LLM by adjusting the probability distribution of tokens~\cite{peeperkorn2024temperature}. A temperature greater than 1 reduces the likelihood of highly probable tokens and enhances the likelihood of less probable tokens. A temperature less than 1 increases the likelihood of highly probable tokens while decreasing the likelihood of less probable tokens. The temperature ranges between 0 and 2, with zero representing a greedy decoding strategy. The default temperature for the Code Llama LLMs is set at 0.2. We did not adjust this value, which clarifies how the output of the Code Llama LLMs is the same across the 10 different runs. 

\subsection{Limitations}
Our work uses prompts two LLMs on coding problems to analyse the effect on energy consumption. Our AI-generated code did not consistently outperform the human-generated code, as seen in related work~\cite{cursaru2024controlled}. Our setup is naive, with one prompt and no hyperparameter tuning. A limitation is that we cannot rule out that different settings would lead to the same results. Next to that, we used a laptop for our measurements. A different hardware setup could lead to different conclusions, e.g. we might see a more distinct deviation if we use a GPU cluster instead. This hardware limitation is reflected in the deviations in Table~\ref{tab:base_code_prompt_against_optimised_code_prompts_all_problems} where the same outcome (similarity 100) sometimes leads to lower or higher energy consumption. Even though we tried to kill background processes and run the device in 'airplane' mode, we could not rule out noise between -4.9\% and 0.5\% in the mean energy measurements.

\subsection{Future Work}
\noindent
\underline{RD1: Hardware energy measurement} to measure more accurately. We also want to test the energy consumption of code on other hardware systems to find out if these have less noise, i.e. microcontrollers or embedded systems.
\\
\hspace{5mm}
\underline{RD2: Expand scope} to include LLMs that are now unavailable such as ChatGPT. Next to that, extend the scope with more programming languages such as C, Java, PHP etc. 

\section{Conclusions} \label{sec:conclusion}
This study sets out to explore to what extent LLMs can generates code with lower energy consumption than human-written code. We conduct experiments to examine the impact of prompt modification on the energy consumption of code generated by LLMs. We use several code-level energy-efficient best practices, popular code LLMs and code problems.

The answer to our main research question \textbf{To what extent can prompts trigger a LLM to produce code that consumes less energy?} is quite nuanced. We find that the \texttt{for-loop} prompt optimisation often results in a lower energy consumption compared to the base solution. However, the instructions provided by the \texttt{for-loop} prompt optimisation are not always implemented correctly. Some LLM-generated solutions outperform the base ones due to the replacement of native Python code with more efficient library functions. A poignant counterexample consists of a \texttt{for-loop} optimisation leading to a 465.5\% increase in energy consumption compared to the base solution. 
The \texttt{for-loop} optimisation also does not consistently reduce energy consumption across all LLMs. 

\section*{Acknowledgments} 
We thank Oracle, in particular Bas Oudejans and Marc Ordelman for their support. This research was supported by GreenDIGIT, an European Union project funded under the grant agreement 101131207, and the Universiteit van Amsterdam Master Artificial Intelligence.
\bibliography{references}
\bibliographystyle{IEEEtran}
\end{document}